# Enzyme as a thermal resonance pump

Korotkov EV

*Bioengineering Center of Russian Academy of Sciences, Prospect 60-tya Oktyabrya, 7/1, 117312, Moscow, Russia.*

Email: bioinf@yandex.ru

fax:7-095-135-0571
tel:7-095-135-2161

**We found latent periodicity of 150 protein families now. We suppose that latent periodicity can determine a spectrum of resonance oscillations in proteins.** Method of Information Decomposition (ID) of symbolical sequences permitted to reveal the latent periodicity of amino acid sequences (AAS)[1-4]. Latent periodicity (LP) found by the ID method was used as the primary profile for search of the LP in protein families with the possibility of deletions and insertions of the symbols in AAS. Swiss-Prot data bank was scanned by the primary profile with using of the modified profile analysis[5] for search of AAS that have a statistical important level (score>6.0) of cyclic alignment against the primary profile[2,3]. After this scanning the profile was recalculated with using of found cyclic alignments that belong to the protein family where the primary profile was detected. At the same time cyclic alignments belong to other protein families were considered as a noise[2,3]. The recalculated profile was used for the new scanning of Swiss-Prot data bank. New statistical important AAS with cyclic alignment against the recalculated profile were not revealed after 5-7 iterations. And as result the LP was revealed for more than 80% of proteins from the protein family where the primary profile was detected firstly. At the present time different types of the LP were found in the 150 protein families, such as protein kinases, cytochromes, NAD+ binding sites. These protein families contain more than 6 thousands of proteins from Swiss-Prot data bank. Full list of the proteins families can be received from web sites http://bioinf.narod.ru/periodicity100 and http://bioinf.narod.ru/periodicity/new. Fig shows the example of the LP for NAD+ binding site. Received results show that LP is common property for many protein families having different biological functions. Different protein families have different types[1] of the LP.

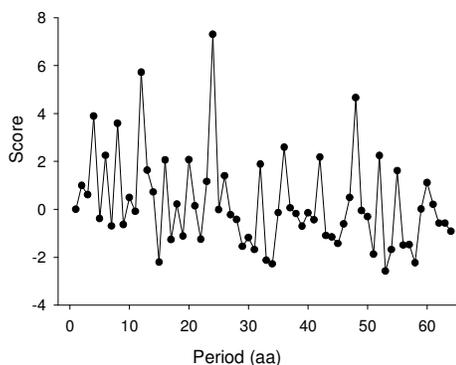

Fig. Information decomposition of the NAD+ binding site. Sequence from 10 to 151 amino acid (accession number is P22414) was aligned against the LP with period equals to 24 amino acids. Alignment is shown at site http://bioinf.narod.ru/periodicity/fam1.txt.

It is very interesting to discuss the possible biological meaning of the LP of AAS. It is possible to suppose that LP can reflect very ancient processes of gene or protein origin by numerous tandem duplications. Or it is possible that the LP could be related with domain or spatial organization of proteins.[1-4] However a conservation of the periodicity in AAS for long time shows that LP may correspond to some function. It is possible to assume that each type of LP can create an own frequency spectrum of oscillations in AAS. This spectrum should depend on the period length and the type of amino acids which create the LP. Thermal moving of molecules of water may excite oscillations exactly on these own frequencies of a protein. The spectrum of these frequencies should be specific for biological functions carrying out by a protein. If a protein carries out the certain enzyme reaction than own resonance frequencies of a protein should be specific for substrate transformations. Discovery of the LP of protein families may evidences that a considerable part of known proteins may work as a resonance molecular machine that uses pumping of thermal energy in own resonance frequencies.